# Spin fluctuations and ferromagnetic order in two-dimensional itinerant systems with Van Hove singularities

P. A. Igoshev[a*], A. A. Katanin[a,b], H. Yamase[b], V. Yu. Irkhin[a]

[a]*Institute of Metal Physics, Kovalevsakya str,18., Ekaterinburg, 620041, Russia*
[b]*Max-Plank-Institute for Solid State Research, Stuttgart, 70569, Germany*

**Abstract**

The quasistatic approach is used to analyze the criterion of ferromagnetism for two-dimensional (2D) systems with the Fermi level near Van Hove (VH) singularities of the electron spectrum. It is shown that the spectrum of spin excitations (paramagnons) is positively defined when the interaction between electrons and paramagnons, determined by the Hubbard on-site repulsion $U$, is sufficiently large. Due to incommensurate spin fluctuations near the ferromagnetic quantum phase transition, the critical interaction $U_c$ remains finite at VH filling and exceeds considerably its value obtained from the Stoner criterion. A comparison with the functional renormalization group results and mean-field approximation which yields a phase separation is also performed.
© 2008 Elsevier B.V. All rights reserved

## Introduction

Layered magnetic systems with strong electronic correlations have been attracting attention of researchers for more than two decades. The compound $UGe_2$ is ferromagnetic at low temperatures and under pressure. It is expected that the ferromagnetic fluctuations may play a crucial role in the properties of the paramagnetic compound $Sr_2RuO_4$ [1]. This is confirmed by the properties of the La-doped compound $La_xSr_{2-x}RuO_4$ which is near the ferromagnetic transition at x=0.27 [2], and also by the properties of the isoelectronic compound $Ca_2RuO_4$ [3], which becomes ferromagnetic under pressure. The above-mentioned compounds have a Van Hove singularity (VHS) of the density of states near the Fermi level.

According to the Stoner theory, the large value of the density of states at the Fermi level, that occurs due to VHS, leads to a possibility of ferromagnetically ordered ground state. At the same time, the Stoner theory does not explain correctly magnetic and thermodynamic properties. In particular, it predicts very high temperatures of magnetic phase transition in comparison with experimental data.

---

[*] P. A. Igoshev.
*E-mail*: igoshev_pa@imp.uran.ru

It is well known [4] that dominant contribution to thermodynamic properties of weak ferro- and antiferromagnets above the Curie temperature comes from diffusive magnetic excitations (paramagnons). Paramagnons are magnetic collective excitation similar to magnons, but in the disordered phase. In the random phase approximation (RPA) paramagnons are non-interacting, but have a strong damping owing to decay of collective excitations into one-particle ones. The attempt of improving RPA is the Hertz-Moria-Millis theory [4-5] which takes into account the paramagnon interaction as a perturbation. In the presence of VHS, RPA and its generalization (like the Hertz-Moriya-Millis theory) are not applicable to describe the ground-state ferromagnetic (FM) transition. Indeed, even for curved Fermi surface flattened sheets of Fermi surface, which provide large density of states on the Fermi level, favour at the same time a spin density wave instability if we do not take into account correlation effects.

An attempt to go beyond these theories was made by Hertz and Klenin [6], who considered the sum of infinite series of diagrams for polarization operator. However, the renormalization of the momentum dependence of the susceptibility, which becomes crucial in the VHS situation, was not performed in their study. The aim of the present paper is to investigate the possibility of FM instability in the presence of VHS.

It is argued that for large enough electron-electron interaction $U_c$ the spectrum of paramagnons becomes positively defined, which corresponds to the possibility of ferromagnetism at $U > U_c$. The quantity $U_c$ exceeds substantially the corresponding value, obtained from the Stoner criterion. The latter is in agreement with the results of the functional renormalization group approach [7] and numerical investigations for the Hubbard model.

### Applicability of the Stoner criterion

To investigate the effect of VHS on magnetic properties of 2D systems we use the mapping of the Hubbard model onto the spin-fermion model [8], which is valid in the vicinity of magnetic phase transition, the coupling constant $U$ for both the models being equal. The generating functional of the spin-fermion model has the form

$$Z_{sf}[\eta,\eta^+] = \int D[c,c^+;\mathbf{S}]\exp[-\beta S],$$
$$S = \sum_{k\sigma} c^+_{k\sigma}(-i\nu_n + \varepsilon_\mathbf{k})c_{k\sigma} + U\sum_q \mathbf{S}_q \mathbf{s}_{-q} + S_S$$
$$\quad -\sum_{k\sigma}(\eta^+_{k\sigma}c_{k\sigma} + c^+_{k\sigma}\eta_{k\sigma}), \quad (1)$$
$$S_S = \sum_q \chi_q^{-1} \mathbf{S}_q \mathbf{S}_{-q},$$

where the Grassman fields $c_{k\sigma}, c^+_{k\sigma}$ correspond to electronic degrees of freedom, the field $\mathbf{S}_q$ describes the collective spin degrees of freedom, corresponding to paramagnons, $\chi_q$ is the "bare" susceptibility of the paramagnon subsystem (to keep the correspondence to the Hubbard model we assume $\chi_q^{-1} = \Pi_q^{-1} - U + U^2\Pi_q$ [8]). $\mathbf{s}_q = \sum_{k\sigma\sigma'} c^+_{k\sigma}\boldsymbol{\sigma} c_{k+q\sigma'}$ is the spin operator of itinerant electrons, $\boldsymbol{\sigma}$ is the Pauli matrix vector, $\Pi_q$ is the exact electronic polarization operator, $\eta,\eta^+$ are the fermionic source fields, $\beta = 1/T$ is the inverse temperature. We use 4-vector notation, e.g., $k=(\mathbf{k};i\nu_n)$, where $\nu_n$ is the fermionic or bosonic Matsubara frequency.

The electronic spectrum has the form
$$\varepsilon_\mathbf{k} = -2t(\cos k_x + \cos k_y) + 4t'(\cos k_x \cos k_y + 1) - \mu \quad (2)$$
where $t$ and $t'$ are the nearest- and next-nearest neighbor hopping parameters, $\mu$ is the chemical potential. For $t' < 0.5t$ the non-interacting density of states has logarithmic singularity at the energy $4t - 8t'$ measured from the bottom of the band. For $t' = 0.5t$ there are VHS lines along the $k_x=0$ and $k_y=0$ directions, and the density of states has stronger divergence, $\rho(\varepsilon) \propto \varepsilon^{-1/2}$ at the bottom of the band (the so-called flat band case, which is analogous to the giant VHS in 3D systems [9]).

The dynamical susceptibility of the non-interacting electronic gas with the dispersion (2) is
$$\chi_q^0 = -T\sum_k G_k^0 G_{k+q}^0, \quad G_k^0 = (i\nu_n - \varepsilon_\mathbf{k})^{-1}, \quad (3)$$

where $G_k^0$ is the bare electronic Green function. In the presence of VHS the uniform susceptibility is logarithmically divergent near Van-Hove band fillings ($\mu = 0$):

$$\chi_0^0 \cong \frac{1}{2\pi t}\ln\left(\frac{t}{\max(|\mu|,T)}\right). \quad (4)$$

According to the Stoner criterion, this leads to a possibility of ferromagnetism for $U > U_c^{\text{Stoner}} = 1/\chi_0^0$. However, the momentum dependence $\chi_{(\mathbf{q};0)}^0$ at low $T \ll t$ and $\mu \neq 0$ has its maximum at $\mathbf{q} \neq 0$. In RPA

$$\chi_q^{\text{RPA}} = \frac{\chi_q^0}{1 - U\chi_q^0}, \qquad (5)$$

this results in an instability of FM ground state, since $\chi_q^{\text{RPA}}$ is divergent at $\mathbf{q} \neq 0$ even for $U < U_c^{\text{Stoner}}$, which implies the spin-density wave instability in this approach. This shortcoming holds also in the Hertz-Moriya-Millis theory [4,5], considering RPA susceptibility with momentum-independent correction in the denominator and neglecting renormalization of momentum dependence of susceptibility. The correlation effects beyond RPA suppress the tendency to incommensurate fluctuations. Therefore there exists some critical value $U_c$ separating paramagnetic (PM) phase for $U < U_c$ and FM phase for $U > U_c$, provided that Fermi level lies in the vicinity of VHS. Determining the phase boundary of PM and FM phases is therefore a delicate problem which requires taking into account all orders of perturbation theory.

**Magnetic susceptibility**

To investigate magnetic properties of the model (1) we consider an exact magnetic susceptibility, defined as the causal double-time Green function

$$\chi_q^{\text{el}} = \frac{1}{4}\int_0^\beta d\tau e^{i\omega_n \tau}\langle T_\tau s_\mathbf{q}^+(\tau) s_{-\mathbf{q}}^-(0)\rangle \qquad (6)$$

and the exact paramagnon propagator

$$R_q = \int_0^\beta d\tau e^{i\omega_n \tau}\langle T_\tau S_\mathbf{q}^+(\tau) S_{-\mathbf{q}}^-(0)\rangle \qquad (7)$$

Here $s_q^\pm(\tau)$ and $S_q^\pm(\tau)$ are the operators in the Heisenberg representation, $T_\tau$ denotes the imaginary-time ordered product. It can be shown that $\chi_q^{\text{el}} = \Pi_q / (1 - U\Pi_q)$. This relation can be obtained from RPA by dressing of non-interacting susceptibility, which results in $\chi_q^0 \to \Pi_q$. On the other hand, if we take into account the self-energy corrections to the bare paramagnon propagator $\chi_q$, which dramatically renormalize its momentum dependence, we obtain

$$R_q^{-1} = \chi_q^{-1} - U^2 \Pi_q. \qquad (8)$$

The polarization operator $\Pi_q$ can be represented as a set of diagrams, which contain the loops of the electronic Green functions, connected by two or more paramagnon lines. In the present paper we neglect contributions to $\Pi_q$ which contain more than one electronic loop, since they are expected to give small corrections. Then we have

$$\Pi_q = \langle \pi_q[\mathbf{S}]\rangle,$$
$$\pi_q[\mathbf{S}] = -\frac{T}{2}\sum_{k_1,k_2}\text{Tr}_\sigma[G_{k_1,k_2}[\mathbf{S}]\sigma^z G_{k_2+q,k_1+q}[\mathbf{S}]\sigma^z] \qquad (9)$$

According to the Mermin-Wagner theorem [10], the long-range FM order in 2D systems is possible at $T = 0$ only. The necessary condition of the FM ground state is

$$U^2 \chi_0 \Pi_{q=0} \to 1 \text{ at } T \to 0 \qquad (10)$$

(generalized Stoner criterion), and the positivity of the spectrum of static paramagnons $\omega_\mathbf{q} = R_{(\mathbf{q};0)}^{-1}$ at $T=0$, which is fulfilled if $\Pi_{(\mathbf{q};0)}$ is maximal at $\mathbf{q} = 0$. This criterion is violated in RPA where $\omega_\mathbf{q}$ is not positively defined in the $T \to 0$ limit in the lowest (second) order of perturbation theory in $U$. To investigate the possibility of ferromagnetism we consider below the results for the polarization operator $\Pi_q$ beyond RPA.

**The static and quantum contributions**

Under the assumption of ferromagnetically ordered ground state, the considered expressions can be simplified in the $T \to 0$ limit, similar to the earlier investigated case of antiferromagnetic order [11-13]. It can be shown that for $(t/T)^{1/2} \ll \xi$ only the contribution of classical spin fluctuations is important. The latter condition is surely fulfilled in the 2D case at finite low $T$ above the ordered ground state, since the correlation length is exponentially large at small $T$ (cf. [14,15]). This distinguishes the present theory from the 3D case [6] where above-discussed approximations give only qualitatively, but not quantitatively correct description of magnetic properties.

Analogous approximations are also applicable to higher order diagrams. Performing these approximations yields the same results as for the model with the action

$$Z_{\xi\to\infty}[\eta,\eta^+] = \int d\mathbf{S}\exp[-3U^2\mathbf{S}^2/(2\Delta^2) \\ -\beta\sum_k \eta_k^+(i\omega_n - \varepsilon_\mathbf{k} - U\boldsymbol{\sigma}\mathbf{S})^{-1}\eta_k] \quad (11)$$

which contains only one uniform static mode **S** with the propagator $\Delta^2 = 3TU^2\sum_\mathbf{q}\omega_\mathbf{q}^{-1}$. The Green functions of the Bose and Fermi fields in the functional (11) can be expressed in a closed analytical form, cf. [12,13]. The static approximation and neglecting **q**-dependence of electronic Green function lead to zero momentum and frequency transfers along the paramagnon lines in all the diagrams. Then $G_{kk'}[\mathbf{S}]$ becomes diagonal with respect to quasimomenta and frequency:

$$G_{kk'}[\mathbf{S}] \to G_k(\mathbf{S})\delta_{kk'}, G_k(\mathbf{S}) = (i\omega_n - \varepsilon_\mathbf{k} - U\boldsymbol{\sigma}\mathbf{S})^{-1}. \quad (12)$$

Therefore we obtain

$$\Pi_q = \langle \pi_q(\mathbf{S})\rangle_{\xi\to\infty}, \quad (13)$$

where $\pi_q(\mathbf{S})$ is the polarization operator of free electrons in the external magnetic field **S**, the subscript $\xi\to\infty$ corresponds to averaging with the functional (11). Similarly we find the electronic Green function: $G_k = \langle G_k(\mathbf{S})\rangle_{\xi\to\infty}$. An explicit expression for $G_k$ was obtained earlier in the paper [13]. The corresponding spectral function

$$A(\mathbf{k},\omega) = \frac{9}{\sqrt{6\pi}\Delta^3}(\omega - \varepsilon_\mathbf{k})^2 \exp\left[-\frac{3(\omega - \varepsilon_\mathbf{k})^2}{2\Delta^2}\right] \quad (14)$$

has a two-peak (non-quasiparticle) structure at the Fermi surface, which destroys the quasiparticle picture due to strong FM fluctuations. As discussed in the paper [15], the corresponding violation of Fermi-liquid behavior corresponds to a quasi-splitting of the Fermi surface at low temperatures, which is related to the change of the electronic spectrum in the vicinity of the magnetically ordered ground state.

### Magnetic Phase Diagram

To investigate the criterion of the FM ordering in the ground state we consider the momentum dependence of the static polarization operator $\Pi_{(\mathbf{q};0)}$ at $T\to 0$. $\Pi_{(\mathbf{q};0)}$ was calculated according to the Eq. (13) for $t'=0.45t$ and different values of $n$ and $\Delta$. The chemical potential $\mu$ is adjusted to keep the number of electrons $n = 2\sum_\mathbf{k}\int_{-\infty}^{\mu}d\varepsilon A(\mathbf{k},\varepsilon)$ equal to the non-interacting value with increasing $\Delta$, the spectral function $A(\mathbf{k},\varepsilon)$ being determined by the Eq. (14).

In the case $\Delta=0$ (i.e. $U=0$), irreducible spin susceptibility coincides with (3) and has no global maximum at **q**=0, but acquires it with increasing $\Delta$: for $n<n_{VH}$ the local maximum at **q**=0 becomes the global one and for $n>n_{VH}$ this maximum shifts to the point **q**=0. Therefore, in both the cases the electron-paramagnon interaction leads to the global maximum of polarization operator of the electronic subsystem at **q**=0, which means a possibility of FM ground state. To determine the critical value of $\Delta$ (the minimum value $\Delta_c$ at which the ferromagnetic ground state is possible) we consider the second derivative of the polarization operator with respect to $q_x$ (or $q_y$). Changing sign of this second derivative determines the critical value $\Delta_c £ t$, which depends on the electronic concentration.

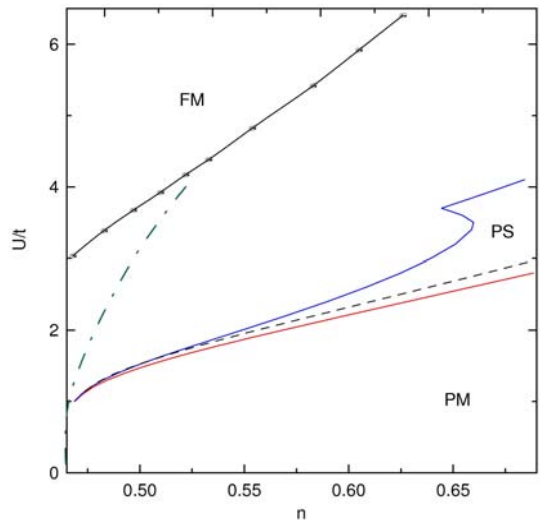

Fig 1: Ground state phase diagram in the *U-n* plane for t'/t=0.45, $n>n_{VH}$ = 0.46. Solid line with dots is the phase boundary between paramagnetic (PM) and ferromagnetic (FM) states in the quasistatic approximation, dot-dashed line is the fRG result [7], solid lines are the boundaries of FM (upper line) and PM (lower line) phases in the mean-filed approach ($T=0.001t$) with phase-separation (PS) region between them, dashed line is the Stoner criterion result

To establish the relation between the critical value $\Delta_c$ and the corresponding value of the Hubbard interaction $U_c$ we use the generalized Stoner criterion (10). For the Hubbard model we find $U_c=1/\Pi_0(\Delta_c)$ where $\Pi_0$ is the uniform static polarization operator. The resulting phase diagram in the $n$-$U$ plane for $t'=0.45t$ and $n>n_{VH}$ is presented in Fig. 1. For comparison, we also show the values $U_c^{Stoner}$ obtained from the standard Stoner criterion and the phase boundaries of the ferro- and paramagnetic state obtained in the mean-field approximation. Similar to the previously studied Pomeranchuk instability [16], the mean-field approximation yields a first-order transition from para- to ferromagnetic state with a rather narrow phase-separation region between them.

The critical values $U_c$ in the quasistatic approximation are larger than $U_c^{Stoner}$. The results for $n$ far from $n_{VH}$ are in quantitive agreement with the functional renormalization group (fRG) results for the Hubbard model [7]. However, in contrast to fRG results, the critical value $U_c$ is non-zero for $n=n_{VH}$, which is related to the breakdown of the quasiparticle picture of the electronic spectrum owing to FM fluctuations. In the case of a flat band ($t'/t=0.50$) the phase diagram does not change so dramatically in comparison with the mean-field theory.

### Conclusions

We have considered the necessary condition for the existence of ferromagnetism in 2D systems with VHS: the maximum of the polarization operator at **q**=0, which ensures that magnetic excitation spectrum is positively defined. We have shown that with increasing $U$ this condition is fulfilled. The critical value of the electron-electron interaction $U_c$ for which the ferromagnetic ground state is possible exceeds substantially the corresponding values, determined from the mean-field theory, and agrees with the fRG investigations of the Hubbard model for the electronic concentration away from Van Hove filling. The critical values $U_c$ are finite at the Van Hove electronic filling for $t'/t<0.50$, which is related to the non-quasiparticle picture of the electronic spectrum (quasi-splitting of the Fermi surface [15]). Thus, the non-Fermi-liquid properties of the electronic spectrum in 2D systems become important for the criterion of the ferromagnetism in the vicinity of VHS.

The picture of the transition from the ferro- to paramagnetic phase remains an open question. An existence of an intermediate state with strong short-range order, characterized by the wave vector **Q**≠0 is possible. Another possibility is that the phase separation, obtained in the mean-field approach and not considered in the quasistatic approximation, is more energetically favourable. This situation is similar to that for the 3D nearly half-filled Hubbard model where the phase-separation state (between ferro- and antiferromagnetic phases) turns out to be more stable than spiral spin configurations [17].

The problem of ferromagnetism formation at $n<n_{VH}$ requires further investigation including competition of spin fluctuations at large- and small momenta, so that the change of momentum dependence of susceptibility possibly plays an important role in this case too.

The work was supported in part by RFFI grants 07-02-01264-a and 1941.2008.2.

### Bibliography

1. I. I. Mazin and D. J. Singh, Phys. Rev. Lett. 79 (1997) 773.
2. N. Kikugawa and Y. Maeno, Phys. Rev. B 70 (2004) 134520.
3. F. Nakamura, T. Goko, M. Ito et al., Phys. Rev. B 65 (2002) 220402.
4. T. Moriya, *Spin Fluctuations in Itinerant Electron Magnetism*, Springer: Berlin, 1985.
5. J. A. Hertz, Phys. Rev. B **14**, 1165 (1976);. A. J. Millis, Phys. Rev. B 48 (1993) 7183.
6. J. A. Hertz and M. A. Klenin, Phys. Rev. B 10 (1974) 1084.
7. A. A. Katanin and A. P. Kampf. Phys. Rev. B 68 (2003) 195101.
8. P. A. Igoshev, A. A. Katanin, V. Yu. Irkhin, Sov. Phys. JETP 105 (2007) 1043.
9. S. V. Vonsovskii, M. I. Katsnelson, and A. V. Trefilov, Phys. Met. Metallogr. 76 (1993) 247.
10. N. D. Mermin, H. Wagner, Phys. Rev. Lett. 17 (1966) 1133; T. Koma, H. Tasaki, Phys. Rev. Lett. 68 (1992) 3248.
11. J. Schmallian, D. Pines, B. Stoikovic, Phys. Rev. B. 60 (1999) 667.
12. M. V. Sadovskii, Physics Uspekhi 44 (2001) 515.
13. A. A. Katanin, Phys. Rev. B 72 (2005) 035111.
14. J. Vilk and A.-M. S. Tremblay, J. Phys. I (France) **7** (1997) 1309.
15. A. A. Katanin, A. P. Kampf, and V. Yu. Irkhin, Phys. Rev. B 71 (2005) 085105; A. A. Katanin and V. Yu. Irkhin, Phys. Rev. B 77 (2008) 115129.
16. H. Yamase, V. Oganesyan, and W. Metzner, Phys. Rev. B 72 (2005) 035114.
17. P. B. Visscher, Phys. Rev. B 10 (1974) 932, 943.